\documentclass[12pt,a4paper]{article}
\usepackage{graphicx}
\usepackage[cp1251]{inputenc}
\usepackage[english]{babel}
\usepackage{amssymb, amsfonts, amsmath}
\usepackage{amscd}
\begin{document}

\title{Separation of variables in Hamilton-Jacobi equation for a charged test particle in the Stackel spaces of type (2.1)}

\author{Valeriy Obukhov}

\date{}

\maketitle
\abstract Found all equivalence classes for electromagnetic potentials and space-time metrics of Stackel spaces, provided that the equations of motion of the classical charged test particles are integrated by the method of complete separation of variables in the Hamilton-Jacobi equation. Separation is carried out using the complete sets of mutually-commuting integrals of motion of type (2.1), whereby in a privileged coordinate system the Hamilton-Jacobi equation turns into a parabolic type equation.


Tomsk State Pedagogical University, 60 Kievskaya St., Tomsk, 634041, Russia.

Tomsk State University of Control Systems and Radioelectronics, 40  Lenin Ave.,

Tomsk,  634050, Russia

\section{Introduction}

\qquad The theory of Stackel spaces is one of the consequences of the symmetry theory. A distinctive feature of a Stackel space is the existence of a so-called complete set in it, consisting of Killing vector and tensor fields, mutually commuting, and satisfying some additional conditions. This makes it possible to carry out a complete separation of variables in classical and quantum equations of the test particle motion. (see for example\cite{a1} -\cite{nova4}). Stackel spaces were named after\quad $Paul\quad  St\ddot{a}ckel$,\quad  who found the first example of a similar space \cite{v1}. Besides Stackel, the main contribution to the construction of the theory was made by Levi-Civita \cite{v2}, Yarov-Yarovoy \cite{v3}, and V.V.Shapovalov \cite{v4} -\cite{v5}. Shapovalov proved the basic theorem of the theory of Stackel spaces.  In \cite{v6}, the theory was generalized to the case of complex privileged coordinate systems.

Shapovalov's theorem allowed invariant partitioning of a set of Stackel metrics into equivalence classes.
 According to this partition, the space itself and its complete sets belong to \quad $(N.N_0)$\quad type, where $N$ is the maximum number of independent Killing vector fields \quad ${Y_p}^i$ \quad included in the full set;\quad $ N_0 = N-\det({Y_p} ^i{Y_q}_i) $.\quad It can be said in such cases that there is a complete separation of type \quad $(N.N_0)$. \quad In the case of the Lorentz signature, \quad $N-0=0,1$.\quad If\quad  $N-0=0$,\quad the space is called non-null. Otherwise, null. There are seven disjoint sets of Stackel spaces with the given signature; respectively: four non-null ones: (0.0), (1.0), (2.0), (3.0); and three null ones (3.1), (2.1), (1.1). A more detailed description of the theory and a rather detailed bibliography can be found in the works \cite{nova1}-\cite{nova4}.

Although there are some examples of successful application of non-commutative integration methods for classical and quantum equations of test particles motion  (see \cite{Ncom 1}-\cite{Ncom 4}), the Stackel spaces, thanks to their rich geometric content remain the most interesting object for research in various sections of theoretical and mathematical physics. They occupy a special place in the theory of gravitation(see for example \cite{b1}-\cite{b10}) and cosmology, since the Robertson-Walker space belongs to the Stackel class. A high level of space symmetry allows a successful construction of new cosmological models, particularly those which include dark energy or features of modified theories of gravity  (see for example
\cite{c1}, \cite{c2}), and also allows the use of symmetry theory methods to further justify the choice of models of extended gravity theory (see for example \cite{c3}).

Shapovalov's theorem allows to classify the metrics of the Stackel spaces in the presence of physical fields of different nature. The classification refers to the enumeration of all equivalence classes for metrics and fields with respect to a group of allowable (that is, not violating the conditions of the complete separation of variables) coordinates and potentials transformations. The set of all Stackel spaces is broken down into disjoint subsets containing spaces with metrics of type \quad ${(N.N_0)}$.\quad Therefore, classification is carried out separately for each type of space. A large number of works in the theory of gravitation are devoted to the classification problems. (see, for example, the review in \cite{nova1}).
However, they have not been studied until recently for the case where a charged test particle moves in an external electromagnetic field in the absence of any restrictions on that field. For the first time this classification problem has been set and solved in our works: for the spaces of type (2.0) - in \cite{nova1}; for the spaces of type (1.0) - in \cite{nova2}.
In this paper, the spaces of type $(2.1)$ are studied. Metrics of this type have been investigated in both flat space-time (\cite{pl1}-\cite{pl4}) and curved one  (see for example \cite{(2.1)1} -\cite{(2.1)7})

The Hamilton-Jacobi equation for a charged test particle has the form:
\begin{equation}\label{1} g^{ij}P_iP_j  = \tilde{m},\end{equation}
where \quad $P_i = p_i + A_i = \partial S/\partial u^i +A_i$.\quad Everywhere in the text, lowercase letters with the $~$ $tilda$ icon denote constants.

Note that the paper \cite{0} considered the problem of complete separation of variables in the "natural Hamilton-Jacobi equation" containing in addition to the electromagnetic potential a quite arbitrary scalar field. In this case, there are no additional constraints on the Stackel space metric, and the potential is to be set by the Stackel matrix and arbitrary functions, each of which depends on one of the non-ignorable variables.

The complete integral of the equation \eqref{1} can be represented in the form:
\begin{equation}\label{2} S=\sum{s_{i}(u^i)}.\end{equation}
The coordinate system, in which it is possible is called privileged, and it is denoted by the variables: \quad $u^i, \quad i,j= 0,\dots,3$. \quad Everywhere in the text, there is a summation within the specified limits of index change on repeated upper and lower indexes.
By diagonalizing the Killing vector fields \quad $Y_p ^{i}$ \quad from the full set, let us bring the functions \quad $s_p$  \quad in \eqref{2} to the form: \quad $s_p=\tilde{\lambda}_p u^p$.\quad Therefore \quad $u^p$\quad are ignorable coordinates.  Let us denote their coordinate indices with letters \quad $  p,q = 0,1, \dots = N-1.$ \quad Non-ignorable coordinates will have the  indexes \quad $\mu,\nu = N, \dots,3$.\quad Functions that depend only on the variable \quad $u^2$ \quad will be denoted by the lowercase Greek letters, and only on the variable \quad $u^3$\quad - by the lowercase Latin letters.
Exceptions: Kronecker characters: \quad $\delta^{ij}, \delta_{ij}, \delta^{i}_{j}$ ; \quad components of metric tensor: $g^{ij}, g_ {ij}$  ; \quad  $\varepsilon, \varepsilon_i=+1,-1; \quad \lambda,\lambda_{i} = const $; \quad  $h_{\nu}^{ij}, h_\nu^i, h_\nu$  - function of \quad $u^{\nu}$.
In these notations, the metric tensor of space, vector, and tensor fields of Killing can be written as:
\begin{align}\label{3} g^{ij}=X^{ij}_3 = (\hat{\Phi}^{-1})_3^\nu h^{ij}_\nu =\frac{\delta^i_p\delta^j_q h^{pq} +\delta^i_a{\delta^j_b}\beta^{ab}+\varepsilon^3\delta^i_3\delta^j_3}{\Delta},\cr
Y^i_p=\delta^i_p, \quad X^{ij}_2=\frac{f\delta^i_a\delta^j_b\beta^{ab} -\phi\varepsilon_3\delta^i_p\delta^j_qh^{pq}+- \phi\varepsilon_3\delta^i_3\delta^j_3}{\Delta} \end{align}
$$\hat{(\Phi})^{\mu}_{\nu}=\begin{pmatrix} 1 & \phi\\ -1 & f \end{pmatrix},\quad \Delta = \phi+ f=\det\hat{\Phi},\quad h^{pq}=h^{pq}_3,\quad \beta^{ab}=h^{ab}_2 ,\quad   $$ $a,b=0,1,2$;\quad $ p,q, = 0,1$; \quad $\nu, \mu = 2,3$.

Since Killing vector fields satisfy the condition:
\begin{align}\label{5}\det({Y_p}^i{Y_q}_i)=0  \to \det(g_{pq})=0 \to g^{22}=0, \end{align}
the Hamilton-Jacobi equation is a parabolic type equation.

For the Hamilton-Jacobi free equation, the complete set of integrals of motion has the form:
\begin{equation}\label{5} \hat{X_\nu} = X^{ij}_\nu p_ip_j,\quad \hat{X_q} = Y_q^ip_i,\quad (Y_q^i = \delta_q^i),
\end{equation} 
When the classification is implemented, complete sets of integrals of motion for the Hamilton-Jacobi equation in the external electromagnetic field will be found.
The full integral in the preferred coordinate system can be reduced to the form:
\begin{align}\label{7}S=\lambda_0 u^0 + \lambda_1 u^1 + {s_2}(u^2) + {s_3}(u^3).\end{align}
The complete set of integrals of motion allows you to find the complete integral \eqref{7} as a solution to the system of equations:
\begin{align}\label{8} \hat{X_q} = Y_q^ip_i,\quad \hat{X}_\nu=\lambda_\nu,\quad \lambda_3 = \tilde{m}.\end{align}
Here \quad $\hat{X}_\nu$ \quad integrals of motion in case of a charged particle have the form:
\begin{align}\label{9} \quad \hat{X}_\nu =(\hat{\phi}^{-1})_{\nu}^{\mu}\hat{H}_\mu = (\hat{\phi}^{-1})_{\nu}^{\mu}(h^{ij}_{\mu}p_{i}p_{i}+2h^{i}_{\mu}p_{i}+h_{\mu}),\end{align}
where
$$ h_2^{ij} =\delta^i_a{\delta^j_b}\beta^{ab},\quad h_2^i=\delta^i_a\gamma^a,\quad h_2=\rho,\quad h_3^{ij}=\delta^i_p{\delta^j_q}a^{pq}+\varepsilon_3\delta^i_3\delta^j_3,\quad h_3^i=h^i,\quad h_3=r.$$
The \eqref{9} relationships allow us to find the contravariant components of the metric tensor and the electromagnetic potential, and also to obtain the condition of separation of variables in the equation \eqref{1}. To do this, equate the coefficients before impulses and their products on the right and left in the equation \eqref{9} when $\nu=3$. As a result, we get:
\begin{align}\label{10}A^p=\frac{\gamma^p+h^p}{\Delta}, \quad A^2 = \frac{\sigma}{\Delta}, \quad A^\nu = 0, \quad A_{i}A^i=\frac{\hat{\omega}+\hat{h}}{\Delta}.\end{align}
Using gradient transformations of potential as well as the group of allowable privileged variable transformations:
\begin{align}\label{10}u^p \to \tilde{c}^p_ru^r + \int{\psi^p}du^2,\quad u^2 \to \int{\psi}du^2,\end{align}
bring matrix \quad $\hat{G} = (G^{ab})= (g^{ab}\Delta),$ \quad and \quad $A^b$ \quad to the form:
\begin{align}\label{12}\hat{G}=\begin{pmatrix} a_{0} , & a_{1} , & 1 \\
a_{1} , & a + \beta, & -\phi \\
1, & -\phi, & 0 \end{pmatrix}; \quad  A^0=\frac{h^0}{\Delta}, \quad A^1=\frac{h^1+\gamma}{\Delta}, \quad A^2=\frac{\sigma}{\Delta}, \quad A^3 = 0.\end{align}
Here and further we denote: $\quad a^{00}=a_0,\quad a^{10}=a_1, \quad a^{11}=a, \quad \beta^{11}=\beta$.\quad Functions \quad $\beta^{00}$ \quad and $\quad \beta^{01}$ \quad can become null by the transformation \eqref{10}.
The functions \quad $a_0,\quad a_1,\quad a \quad $ are linearly dependent. In the future, we will need to classify the $\hat{A}=(a^{pq})$ \quad matrices relative to the group \eqref{10}. Let us list all the non-equivalent classes of \quad $\hat{A}$ \quad matrices.
\begin{equation}\label{12}
\left\{\begin{array}{ll}
\hat{A}_{(1)}=\begin{pmatrix} a_0 & a_1 \\ a_1 & a \end{pmatrix},\quad\cr
\hat{A}_{(2)}=\begin{pmatrix} a_0 & a_1 \\ a_1 & -a_0 \end{pmatrix},\cr
\hat{A}_{(3)}=\begin{pmatrix} a_0 & a_1 \\ a_1 & 0 \end{pmatrix},\quad\cr
\hat{A}_{(4)}=\begin{pmatrix} a_0 & 0 \\ 0 & a \end{pmatrix},\quad\cr
\hat{A}_{(5)}=\begin{pmatrix} a & 0 \\ 0 &\varepsilon a \end{pmatrix},\varepsilon=1,-1,\cr
\hat{A}_{(6)}=\begin{pmatrix} a & 0 \\ 0 & 0\end{pmatrix}.
\\\end{array}\ \right.\end{equation}
We'll make another point. When considering the resulting conditions of the complete separation of variables in the equation \eqref{1}, the ratio arises:
$$\det{a^{pq}}= \tilde{a}+\tilde{c}_{pq}a^{pq}.$$
Using valid \eqref{10} coordinate transformations, it can be simplified and represented as:
 $$ \det{a^{pq}}= \tilde{a} $$ - for all matrices  $\hat{A}_{(\alpha)}$ \quad except \quad $\hat{A}_{(3)}.$ \quad Wheh \quad $\alpha = 3$ \quad  we have: $$-\det{a^{pq}}= \tilde{a}+\varepsilon a_0.$$

In the accepted notations the necessary and sufficient condition for the complete separation of variables in the equation \eqref{1}:\quad  $$A^iA_i = \frac{\rho + r}{\Delta}$$ - can be transformed to:
\begin{align}\label{13} (\rho+r+a_{0}\sigma^2-2\sigma h^0)(a_{0}\phi^2 +2\phi a_{1}+a+\beta )=& \cr(h^0\phi+h^1+\gamma-(a_{0}\phi+a_{1})\sigma)^2.\end{align}
Let us present the covariant components of the metric tensor and electromagnetic potential in the following convenient form for calculations:
\begin{align}\label{14}\hat{G}^{-1}=(G_{ab})=(\frac{g_{ab}}{\Delta}) = \begin{pmatrix}
\frac{\phi^2}{G} , & \frac{\phi}{G}, \quad & 1-\frac{\phi B}{G} \\ \frac{\phi}{G} , & \frac{1}{G}, &  -\frac{B}{G} ,\\
1-\frac{\phi B}{G} , & -\frac{B}{G}, & \frac{B^2}{G}-a^{0}\end{pmatrix},\end{align}
\begin{align}\label{15} A_0=\phi A_1 + \sigma, \quad  A_1 =\frac{h^0 \phi +h^1 + \gamma - B\sigma}{G}, \quad A_2=h^0-a_{0}\sigma -B A_1,\quad A_3=0. \end{align}
It is denoted here by:
\quad $G=-(a_{0}\phi_2 +2\phi a_{1}+a+\beta), \quad B=a_{0}\phi +a_{1}$,

Solving the functional equation \eqref{13} is equivalent to solving two overcrowded systems of the second and third degree algebraic equations. The first system includes functions only from the variable \quad  $u^2$,\quad  the second only from the variable \quad  $u^3$. \quad   As already noted, we will look for solutions in the presence of additional symmetry, for which we will require that the electromagnetic field be external. We call the electromagnetic field external if the potentials contain at least one function independent of the metric tensor. We call such functions free. All other functions included in the electromagnetic potential, we will call related. We also believe that the functions included in \eqref{13} are continuous and smooth, therefore there are points on coordinate axes in the neighbourhood of which none of the non-zero functions vanish. In the future, if it is necessary to fix the variables $u^\nu $ in the process of solving functional equations,  we will use exactly such points.

\maketitle

\section{Solutions for the case when at least one of the functions $\mathbf{h^p}$ is free.}

 In this section we find all the solutions for the case when free functions depending on \quad $u^3$ \quad exist. Everywhere in the section, all functions from the variable \quad $u^3$ \quad except \quad $h^p$ \quad are related.

$\mathbf{I}.$ Both \quad$ h^q $\quad functions are free.

Let us use the above-mentioned smoothness condition and consider the equation \eqref{13} at the fixed point \quad$\tilde{u}^2$\quad on the coordinate axis \quad($u^2$).\quad Since \quad $G \ne 0 ,\quad $ the function \quad $ r $ \quad is expressed through the \quad $h^{p}$ \quad functions as follows:
\begin{align}\label{16} r=r_0+\tilde{l} h^0 + l({h}^1 + \tilde{c}{h}^0 + c)^2,\end{align}
Substituting \eqref{16} into \eqref{13} one obtains the equation:
$$ b_{pq}h^p h^q + b_p h^p +b = 0.$$
As\quad $\partial{b_{pq}}/ \partial{h^r}=\partial{b_p}/ \partial{h^r}=\partial{b}/ \partial{h^r}= 0  \to b_{pq}, b_p, b =0.$
Using the transformation group \eqref{10}, one gets:
$$ G=t, \quad \phi=t_1 G, \quad \phi^2=t_0G \quad  \to \quad \phi=\beta=0.$$
$$\gamma-\sigma a^{0}=b_0, \quad \sigma=b_1 \quad \to \quad \gamma=\sigma=0. $$
We obtain the solution to the equation \eqref{13}:
\begin{align}\label{17} \hat{G}=\begin{pmatrix} a_{0}
& a_{1} & 1 \\ a_{1} & a & 0 \\ 1 & 0 & 0 \end{pmatrix} \quad A^p=\frac{h^p}{\Delta},\quad A^2 = A^3 = 0.\end{align}

$\mathbf{II}$. $h^p$ contain one free function.

Substituting \eqref{16} into \eqref{13}, fixing the variable $u^2$ in another point of the coordinate axis and  solving the resulting equation with respect to $h^1$, one gets:
\begin{align}\label{18} h^1=wh^0+u+Q, \quad Q^2 = p{h^0}^2 +2qh^0 +l.\end{align}


Substituting (18) into (13) one obtains the equation:

$$ b_{pq}h^p h^q + b_p h^p + b = 0.$$

As before from \quad $\partial{b_{pq}}/ \partial{h^r}=\partial{b_p}/ \partial{h^r}=\partial{b}/ \partial{h^r}= 0$ \quad it follows: \quad $b_{pq}, b_p, b =0.$ \quad
Let us denote $z=q^2-pl$ \quad and consider separately the variants \quad $z=0$ \quad and \quad $z \ne 0$.  As a result, using the transformation group  (10) , we get:

$\mathbf{A})$ $\mathbf{z}$ $\mathbf{= 0}.$\quad
In this case, \eqref{13} contains summands that have four linearly independent functions $\quad h^0: \quad h^0,\quad {h^0}^2,\quad Q,\quad h^0 Q \quad$, as well as a free member. By denoting:
$$r=r_1+ t{h^0}^2+2nh^0+2m_0 h^0 Q + 2r_0Q,$$
we get the following systems of equations:
\begin{equation}\label{19}
\left\{\begin{array}{ll}
(\phi+w)^2 +p=tG,\cr
\phi+w=m_0 G. \\\end{array}\ \right.\end{equation}
\begin{equation}\label{20}
\left\{\begin{array}{ll}
q+(\phi+w)(\gamma-B\sigma+u)=(n-\sigma)G, \quad \cr
u+\gamma-B\sigma=r_0G,\cr
(u+\gamma -B\sigma)^2+l =(\rho + r_1 + a_{0}\sigma^2) G. \\\end{array}\ \right.\end{equation}
One can show that from the \eqref{19}  system it follows \quad $\phi=0$.\quad Then it's obvious that $\quad a=0.\quad $ Otherwise we get the previous solution \eqref{17}.  The system of equations \eqref{19}, \eqref{20} has a singular non-trivial solution:
$$a_{0}= a_1=a=\gamma=t=m_0=w=p=n=r_0=u=0,\quad q=\tilde{q}, \quad \sigma=-\frac{\tilde{q}}{\beta},\quad h^0=\frac{{h^1}^2}{2q}.$$
We will write it as:
\begin{align}\label{21} \hat{G}=\begin{pmatrix} 0, & 0, & 1 \\ 0, & \beta, & 0 \\ 1, & 0, & 0 \end{pmatrix}; \quad A^0=\frac{h^2 \tilde{a}}{\Delta},\quad  A^1=\frac{h}{\Delta},\quad A^2 =-\frac{1}{2\tilde{a}\beta \Delta}, \quad  A^3 = 0.\end{align}

$\mathbf{B)}$ $\mathbf{z}$ $\mathbf{\ne 0}$.\quad In this case, from the ratio \eqref{18} it follows:
\begin{align}\label{22}  h^p=b^p h+ h^p _0,\end{align}
where\quad $h$ \quad is a free function. \quad Substitute \eqref{22} into \eqref{13}, and equate to zero the coefficient before the function \quad $h^2$.\quad As a result, we get:
\begin{align}\label{23} (\phi b^0-b^1)^2 = \tilde{e}(\phi^2 a_0+2\phi a_1+a+\beta), \quad \tilde{e}=0,1. \end{align}
First consider the variant:

$\mathbf{1)}$ $\tilde{e}=0$. From \eqref{23} it follows \quad $b^1=\phi=0.$ \quad With this taken into account from \eqref{13} we get:
\begin{align}\label{24}  (\gamma+h^1)^2=(\rho+r)(a+\beta),\quad \sigma=0. \end{align}
Non-trivial solutions of the equation \eqref{24} are:
$$ a)\quad  h^1=\gamma=0,\quad  b)\quad a=h^1=0.$$
The contravariant components of the metric tensor and potential have the form:
\begin{align}\label{25} a)\quad \hat{g}=\begin{pmatrix} \frac{a_0}{\Delta} & \frac{a_1}{\Delta} & \frac{1}{\Delta} & 0 \\ \frac{a_1}{\Delta} & \frac{a+\beta}{\Delta} & 0 & 0\\ \frac{1}{\Delta} & 0 & 0 & 0 \\ 0 & 0& 0 &  \frac{\varepsilon}{\Delta}\end{pmatrix}; \quad A^0=\frac{h}{\Delta},\quad  A^1= A^2 = A^3 = 0.\end{align}
\begin{align}\label{26} b)\quad \hat{g}=\begin{pmatrix} \frac{a_0}{\Delta} & \frac{a_1}{\Delta} & \frac{1}{\Delta} & 0 \\ \frac{a_1}{\Delta} & \frac{\beta}{\Delta} & 0 & 0\\ \frac{1}{\Delta} & 0 & 0 & 0 \\ 0 & 0& 0 & \frac{\varepsilon }{\Delta} \end{pmatrix}; \quad A^0=\frac{h}{\Delta}, \quad A^1=\frac{\gamma}{\Delta}, \quad  A^2 = A^3 = 0.\end{align}
$\mathbf{2)}$ $\tilde{e}=1.$ \quad
Let us consider the equation \eqref{23} in the fixed point \quad $u^3= \tilde{u}^3$.\quad As a result, we get: $$\quad \beta=\tilde{a_0}\phi^2 +2\tilde{a}_{1}\phi+\tilde{a},$$ which relative to the \eqref{10} transformations is equivalent to \quad $\beta=0$. The equation \eqref{23} takes the form:
\begin{align}\label{27}(a_0 -{b^0}^2)\phi^2 + 2(a_1 -{b^0}{b^1})\phi +a - {b^1}^2 =0, \end{align}
and has two solutions:

$\mathbf{a})$ \quad $ a_0 ={b^0}^2, \quad a_1 ={b^0}{b^1}, \quad a = {b^1}^2.$ \quad From the equal-zero in the equation \eqref{13} of the coefficient before the function $h$ it follows:
$$h_0^1+h_0^0\phi + \gamma = t(b^1 + b^0 \phi).$$
Solutions are equivalent to the following: \quad $\gamma=h^p_0=0.\quad$
That is why contravariant components of the metric tensor and potential have the form:
\begin{align}\label{28} \quad (g^{ij})=\begin{pmatrix} \frac{{b^0} ^2}{\Delta} & \frac{b^0 b^1}{\Delta} & \frac{1}{\Delta} & 0 \\ \frac{b^0 b^1}{\Delta} & \frac{{b^1} ^2}{\Delta} & \frac{-\phi}{\Delta} & 0\\ \frac{1}{\Delta} & \frac{-\phi}{\Delta} & 0 & 0 \\ 0 & 0& 0 & \frac{\varepsilon}{\Delta} \end{pmatrix};
\quad A^p=\frac{b^p h}{\Delta}, \quad  A^2 = \frac{\sigma}{\Delta},\quad A^3 = 0.
\end{align}

$\mathbf{b})$ \quad $a = {b^1}^2, \quad \phi=0.$ \quad From the equality to zero of the coefficient before the function $h$  in the equation \eqref {13} it follows:

$$ (b^0 b^1-a_1)\sigma=t-\gamma. $$ We believe that \quad $\sigma\ \ne 0$, \quad because otherwise we will get the already considered option.  Therefore \quad $a_1=b^0 b^1,\quad \gamma=0.$ \quad As\quad $b^1 \ne 0\to h^1_0=0 $.\quad  Taking this into account, the equation \eqref{13} will take the form:
$$ ({b^0}^2 -a_0)= \rho + r -2{h_0^0}\sigma. $$
Therefore \quad $a_0={b^0}^2,\quad h^0_0=0.$ \quad Obtained a particular solution \eqref{28}.

The variables \quad $u^2$\quad and \quad $u^3$ \quad enter the equation unsymmetrically. Therefore, to complete the classification, one must consider separately the case where free functions depend on the \quad $u^2$\quad variable. At the same time, if free functions from \quad $u^3$ \quad arise in the process of the solution, the corresponding variants may not be considered.

\section{Solutions for the case when both functions $\mathbf{\gamma}$ and $\mathbf{\sigma}$ are free.}

Everywhere in the text, all functions from the variable \quad $u^2$ \quad except \quad $\gamma$ \quad and \quad $ \sigma$ \quad are related.  Let both \quad $\gamma$\quad and \quad $\sigma$\quad are free.  Consider the equation \eqref{13} at a fixed point \quad $\tilde {u}^2$.\quad Since $G \ne 0 ,\quad $ the function \quad  $ \rho$ \quad  is expressed through the \quad $\gamma$ \quad and \quad $\sigma$ \quad functions as follows:
\begin{align}\label{29} \rho =\rho_0 +\tau_0\sigma^2+\tau_1\gamma^2 - 2\tau\sigma\gamma -2\mu\sigma+2\nu\gamma.\end{align}
Substitute this expression back into the equation \eqref{13}. After the obvious transformations we will get a system of equations:
\begin{equation}\label{30}
\left\{\begin{array}{ll}
a_0\phi^2+2a_1\phi+a=0 \quad \to \quad G=\beta,\quad\cr
a_0\phi^2+a_1=\beta\tau,\quad\cr
a_0+\tau_0=\beta\tau^2, \quad\cr
h^1+\phi h^0=\beta\nu,\quad\cr
h^0+\mu=\beta\nu\tau,\quad\cr
(h^1+\phi h^0)^2=\beta(\rho_0+r).
\\\end{array}\ \right.\end{equation}
From where immediately follows:
$$a=a_0=a_1=0,\quad h^p=0,$$
The contravariant components of the metric tensor and potential have the form:
\begin{align}\label{31} \hat{g}=\begin{pmatrix} 0 & 0 & \frac{1}{\Delta} & 0 \\ 0 & \frac{\beta}{\Delta} & \frac{-\phi}{\Delta} & 0\\ \frac{1}{\Delta} & \frac{-\phi}{\Delta} & 0 & 0 \\ 0 & 0& 0 & \frac{\varepsilon }{\Delta} \end{pmatrix}; \quad A^0=0, \quad A^1=\frac{\gamma}{\Delta}, \quad  A^2 = \frac{\sigma}{\Delta},\quad A^3 = 0.\end{align}

\section{Solutions for the case of nonlinear dependence between functions $\mathbf{\gamma}$ and $\mathbf{\sigma}$.}

Substituting the expression \eqref{29} into the equation \eqref{13}. As a result, we will get the following functional equation:
\begin{align}\label{32} (r+\rho_0 + \gamma^2\pi+(\tau_0 +a_{0})\sigma^2+ 2\sigma\gamma\tau -2(\mu+h^0)\sigma+2\nu\gamma)G=(H+\gamma-B\sigma)^2,\end{align}
where it is denoted by \quad $H=h^0\phi+h^1$.\quad Let us solve the equation \eqref{32} relative to  $\gamma$.

$\mathbf{1})$ $\pi \ne G.$
\begin{align}\label{33} \gamma = \omega_0\sigma^2 +\gamma_0 + \Sigma, \quad \Sigma^2=\mu_0 \sigma^2 +2\mu_1 \sigma+\mu.\quad \mu_1^2-\mu_1\mu \ne0. \end{align}
Substituting \eqref{33} into \eqref{13} one obtains the equation:
$$ \theta_{0}+\theta_{1}\sigma +\theta_{2}\sigma^2 +\theta_{3}\Sigma  +\theta_{4}\sigma \Sigma =0.$$

As\quad $\partial{\theta_a}/ \partial{\sigma} \ne 0 \to \theta_a=0.$ \quad From this it follows:
\begin{equation}\label{34}
\left\{\begin{array}{ll}
(\omega_0-B)=-\xi_0 G ,\cr
(\omega_0-B)^2+\mu_0=G(\omega +a_0).
\\\end{array}\ \right.\end{equation}

\begin{equation}\label{35}
\left\{\begin{array}{ll}
\gamma_0+H=G\psi_0,\cr
(\omega_0-B)(\gamma_0+H)+\mu_1=(\tau-h^0)G,\cr
(\gamma_0+H)^2+\mu=(r+\rho)G.
\\\end{array}\ \right.\end{equation}
Consider the equations of the system in order, using the classification \eqref{12}.

Let us represent the system \eqref{34} as:
\begin{equation}\label{36}
\left\{\begin{array}{ll}
(\phi a_0 +a_1)-\omega_0=\xi_0(a_{0}\phi_2 +2\phi a_{1}+a+\gamma),\cr
{\omega_0}^2-2(\phi a_0 +a_1)\omega_0+\mu_0=\omega(a_{0}\phi_2 +2\phi a_{1}+a+\gamma ) +a_0 \beta +\det{a^{pq}}. \\\end{array}\ \right.\end{equation}
Consider the system \eqref{36} using matrices forms from \eqref{12}, as well as appropriate ratios for \quad $\det{a^ {pq}}$.

1) $\hat{A}_{(1)}$.   In this case, it is obvious that the system \eqref{36} has no solutions because from the first equation of the system there follows a contradiction, \quad since \quad$\xi_0\to a_1=0$,\quad
which is impossible. By reasoning similarly, it can be shown that the \eqref{36} system is contradictory also for the matrices \quad $\tilde{A}_{(2)}$, \quad $\tilde{A}_{(3)}$\quad and \quad $\tilde{A}_{(5)}$. \quad Let us consider the remaining options.

2) $\hat{A}_{(4)}$. $a$ \quad  and \quad  $a_0$ \quad  are linearly independent functions. That is why it follows from \eqref{36}:
$$\omega=\omega_0=\xi_0=\phi=0, \quad \mu_0=\tilde{a},\quad  a=\frac{\tilde{a}}{a_0}.$$

Substitute these expressions into \eqref{35}. As a result, we will get the system:

\begin{equation}\label{37}
\left\{\begin{array}{ll}
(\gamma_0+h^1)=(a+\beta)\psi_0,\cr
\mu_1=(a+\beta)(\tau-h^0),\cr
(\gamma_0+h^1)^2+\mu=(a+\beta)(r+\rho).
\\\end{array}\ \right.\end{equation}
From the first equation follows $\psi_0=\tilde{\psi}$, which is equivalent to: \quad $\psi_0=0 \to h^1=\gamma_0=0.$ The solution to the second equation is equivalent to the following: \quad $\mu_1=\tau=h^0=\beta=0.$
The contravariant components of the metric tensor and potential have the form:
\begin{align}\label{38} \hat{g}=\begin{pmatrix} \frac{\tilde{a}}{a\Delta} & 0 & \frac{1}{\Delta} & 0 \\ 0 & \frac{a}{\Delta} & 0 & 0\\ \frac{1}{\Delta} & 0 & 0 & 0 \\ 0 & 0& 0 & \frac{\varepsilon }{\Delta} \end{pmatrix}; \quad A^0=0, \quad A^1=\frac{\sqrt{\tilde{a}\sigma^2+\tilde{b}}}{\Delta}, \quad  A^2 = \frac{\sigma}{\Delta},\quad A^3 = 0.\end{align}
3) $\hat{A}_{(6)}$. The solution to the system \eqref{36} has the form:
\begin{align}\label{39} \xi_0=\frac{1}{\phi}, \quad\mu_0=0, \quad \omega_0=-\phi \omega,\quad G=\phi^2(a_0+\omega), \quad B=\phi a_0.\end{align}
By substituting \eqref{39} into \eqref{35}, we find that \quad $\mu_1=0$.\quad  Since from the \eqref {39} it follows that \quad $\mu_0=0$,\quad the condition \eqref{33} is broken.

$\mathbf{2})$ $\pi = G.\quad \to \quad \phi(a_0^2+a_1^2)=0.$

As before the function \quad $\sigma$ \quad can be considered free.  The functions \quad $\gamma$ \quad and \quad $\rho$ \quad are expressed through it as follows:
\begin{equation}\label{40} \left\{\begin{array}{ll} \gamma = \omega_0\sigma +\gamma_0 + \frac{\omega}{\sigma+\sigma_0}, \cr
\rho=\rho_0+(a_0-\tau_0^2)\sigma^2-2(h^0+tau)\sigma-\frac{2\mu_1}{\sigma+\sigma_0}-\frac{\mu_0}{(\sigma+\sigma_0)^2}. \\\end{array}\ \right.\end{equation}
By substituting \eqref {40} into \eqref{13} and equating  coefficients before linearly independent functions to zero, we get the following system of functional equations:
\begin{equation}\label{41}
\left\{\begin{array}{ll}
\omega^2=\beta \mu_0,\cr
(\omega_0-a_1)^2=\beta(\tau +a_0),\cr
(a_1-\omega_0)(\phi h^0+h^1+\gamma_0)=\beta(h^0+\tau),\cr
(\gamma_0+\phi h^0+h^1-\sigma_0(\omega_0-a_1))\omega=\beta\mu_1,\cr
(\gamma_0+\phi h^0+h^1)^2-2(a_1-\omega_0)\omega_0=\beta(r+\rho_0).
\\\end{array}\ \right.\end{equation}
It's easy to show that from \quad $a^{pq}=0$ \quad it follows that \quad  $h^p=0$.\quad Therefore $$a_1 \ne 0 \quad \to \phi=0.$$  By extension of the coordinate \quad $u^1$ \quad  transform \quad  $\beta$\quad to \quad $\varepsilon.$ \quad The remaining equations give the following solution:
$$\omega_0=\gamma_0 =\tau_p=\mu_1=\rho_0=0, \quad \omega=\tilde{c},\quad \sigma_0=-\tilde{e}, \quad h^0=\varepsilon \tilde{e} a_1^2 \quad h^1= \tilde{e} a_1.$$

The contravariant components of the metric tensor and potential have the form:
\begin{align}\label{42} \hat{g}=\begin{pmatrix} \frac{\varepsilon a^2}{\Delta} & \frac{a}{\Delta} & \frac{1}{\Delta} & 0 \\ \frac{a}{\Delta} & \frac{\varepsilon}{\Delta} & 0 & 0\\ \frac{1}{\Delta} & 0 & 0 & 0 \\ 0 & 0& 0 & -\frac{\varepsilon}{\Delta} \end{pmatrix}; \quad A^0=\frac{\varepsilon\tilde{e}a_1^2}{\Delta}, \quad A^1=\frac{(\tilde{c}/(\sigma-\tilde{e})+\tilde{e}a)}{\Delta}, \quad  A^2 =\frac{\sigma}{\Delta},\quad A^3 = 0.\end{align}

\section{Solutions for the case of linear dependence between functions $\mathbf{\gamma}$ and $\mathbf{\sigma}$.}

Due to the unsymmetric entrance of the functions \quad $\gamma$ \quad and\quad $\sigma$ \quad into the electromagnetic potential, in case of their linear dependence, two variants should be considered separately when \quad $\sigma$ \quad is a free function, and \quad $\sigma$ \quad  is a related function. However, it is not difficult to make sure that the last variant does not give any new solutions. Therefore, we consider \quad $\sigma$ \quad as a free function.  In this case, the \quad $\gamma$ \quad and \quad $\rho$ \quad functions can be represented as:

\begin{align}\label{44} \gamma=\gamma_0\sigma+\gamma_1, \quad \rho=\rho_0+\tau_0 \sigma^2 + 2\tau \sigma.\end{align}
By substituting \eqref{44} into \eqref{13}, we get the following system of functional equations:
\begin{align}\label{45} \tilde{a}-\gamma_0^2+2\gamma_0(a_{0}\phi +a_{1})+\tau_0(a_{0}\phi^2 +2\phi a_{1}+a+\beta)+a_{0}\beta=0,\end{align}

\begin{align}\label{46} (\gamma_0-(a_{0}\phi +a_{1}))(\gamma_1+h^0\psi+h^1)=(\tau-h^0)(a_{0}\phi^2 +2\phi a_{1}+a+\beta),\end{align}

\begin{align}\label{47} (\gamma_1+h^0\phi+h^1)^2=(\rho_0 + r),\end{align}
Consider the \eqref{45} equation using the \eqref{12} classification.

$\mathbf{1})$ $\hat{A}_{(1)}$. From the equation \eqref{45} it immediately follows:
$$\tau_0=\gamma_0=\beta=\tilde{a}=0 \quad  \to \quad a^{pq} = \varepsilon b^p b^q.$$
It is easy to show that from \eqref{46} follows: \quad $ h^p=hb^p, \quad \gamma_0=0.$ \quad We receive the solution \eqref{28}.

$\mathbf{2})$  $\tilde{A}_{(2)}$. This variant is also reduced to \eqref{28}. Indeed, from the equation \eqref{45} follows:
$\quad \tau_0=\tilde{b},\quad \beta=\tilde{b}(\phi^2+1), \quad \gamma_0=-\tilde{b}\phi,\quad$ $a_0=\tilde{b}\cos{2b},\quad a_1=\tilde{b}\sin{2b},$ \quad where it is denoted by \quad $\tilde{a}=-\tilde{b}^2$

Let us make the coordinate transformation:
$$u^0 \to u^0-\frac{1}{2}u^2,\quad u^1 \to u^1+\frac{1}{2}\int{\phi}du^2,$$
Then the matrix \quad $(a^{pq})_{(2)}$ \quad is converted as follows:
$$  (a^{pq})_{(2)} \quad \to \begin{pmatrix} \cos{b}^2-\sin{b}^2 & \sin{b}\cos{b} \\ \sin{b}\cos{b}  & \cos{b}^2-\sin{b}^2 \end{pmatrix} \to \quad \det{a^{pq}}=\tilde{a}=0.$$
Therefore \quad $\tau=\gamma_0=\beta=0, \quad $ and the solution is reduced to \eqref{28}.

$\mathbf{3})$  $\hat{A}_{(3)}$. As noted, in this case \quad $ a_0=\varepsilon a_1^2.$\quad From the equation \eqref{45} it follows:

$$\gamma_0=-\tau_0\beta, \quad \tau_0(\beta - \phi\tau_0)=0, \quad \beta=\varepsilon-(2\gamma_0 + \tau_0\phi)\phi. $$
Therefore $\tau_0=\gamma_0=0, \quad \beta=\varepsilon.$ \quad From the equation \eqref{45} we get: \quad $h^0=\varepsilon h^1  \to$ a particular solution to \eqref{28}.

$\mathbf{4})$  $\hat{A}_{(4)}$, $\hat{A}_{(5)}$.  Because of the independence of the functions $\quad a,\quad a_0\quad $ the equation \eqref{45} has no solutions.

$\mathbf{5})$ $\hat{A}_{(6)}$.

$\mathbf{a})$ First we assume: \quad $a=0$.\quad The equation \eqref{45} takes the form:
\begin{align}\label{49} -\gamma_0^2+2\gamma_0 a_{0}\phi +\tau_0(a_{0}\phi^2 +\beta)+a_{0}\beta=0,\end{align}
Therefore: $\quad \beta = \tau_0\phi^2,\quad \gamma_0=-\tau_0\phi, \quad G = \phi^2(a_0 + \tau_0).$ \quad Considering this, we will have the equation
\eqref{46} as the following system:
\begin{align}\label{50} \gamma_1+h^1+\tau\phi=0  \to h^1=0,\quad \gamma_1=-\tau\phi ,\end{align}
\begin{align}\label{51} (\gamma_1+h^0\phi+h^1)^2=\phi^2(\rho_0 + r)(a_{0} +\tau_0).\end{align}
Substitute \eqref{50} into \eqref{51}. As a result we get the solution \quad  $ h^0-\tau = \tilde{e}(a_{0} +\tau_0)$,\quad equivalent to the following: \quad $h^0=\tau=0$.

The contravariant components of the metric tensor and potential have the form:
\begin{align}\label{52} \hat{g}=\begin{pmatrix} \frac{a}{\Delta} & 0 & \frac{1}{\Delta} & 0 \\ 0 & \frac{\phi^2 \beta}{\Delta} & \frac{-\phi}{\Delta} & 0\\ \frac{1}{\Delta} & -\phi & 0 & 0 \\ 0 & 0& 0 & \frac{\varepsilon}{\Delta} \end{pmatrix}; \quad A^0=0, \quad A^1=-\frac{\beta \phi \sigma}{\Delta}, \quad  A^2 =\frac{\sigma}{\Delta},\quad A^3 = 0.\end{align}

$\mathbf{b})$ Let now \quad$a_0=0.$ \quad The equation \eqref{45} takes the form:
$$\gamma_0^2=\tau_0(a+\beta) \to \gamma_0=\tau_0=0.$$
From \eqref{46}, \eqref{47} follows:\quad $ h^0=\tau=0, \quad h^1+\gamma_1= \tilde{e}(a+\beta)$,\quad which is equivalent to the  following:\quad $h^1=\gamma_1= 0$.

The contravariant components of the metric tensor and potential have the form:
\begin{align}\label{52} \hat{g}=\begin{pmatrix} 0 & 0 & \frac{1}{\Delta} & 0 \\ 0 & \frac{a+\beta}{\Delta} & -\frac{\phi}{\Delta} & 0\\ \frac{1}{\Delta} & -\frac{\phi}{\Delta} & 0 & 0 \\ 0 & 0& 0 & \frac{\varepsilon}{\Delta} \end{pmatrix}; \quad A^p=0, \quad  A^2 =\frac{\sigma}{\Delta},\quad A^3 = 0.\end{align}

\section{Conclusion.}

It is obvious that the classification problem under consideration is non-trivial for separating variables of type \quad $(N.N_0)$, \quad if \quad $N<3.$ \quad Hence, to complete the classification we need to consider the only left type \quad $(1.1).$ \quad
Note that the results of this classification in addition to the theoretical can also have an applied value, for example, in the study of the axion field problem. This field associated with an axionic dark matter (see \cite{O0} - \cite{O3} ). In the paper \cite{Balakin} static field configuration in the outer zone of a magnetic star is considered. The configuration is formed by interacting quartet of  external fields: axion field, strong intrinsic magnetic field, and axionically induced electric field.

Therefore, in conclusion, we show the acquired results. For each class, the covariant components of the metric tensor and the electromagnetic potential are given. The functions $s_\nu$ for the complete integral \eqref{7} are provided.

$\mathbf{1})$
 \begin{align}\label{52} ds^2=\frac{1}{ \Delta}[2du^0du^2+\frac{(du^1-a_1du^2)^2}{a} - a_0 {du^2}^2 +{du^3}^2],\end{align}

 $$A_0=0, \quad  A_1 =\frac{h^1}{a}, \quad A_2=h^0,\quad A_3=0. $$

 $$s_2 = \int(\frac{\lambda \psi + \lambda_2 }{2(\lambda_0 - \phi \lambda_1 )})du^2,$$

$$
s^3 = {\varepsilon} \int\sqrt{(   \lambda f -(\lambda_2 + a_0{\lambda_0}^2 +a{\lambda_1}^2 +2a_1 \lambda_0 \lambda_1 +2h^0\lambda_0+2 h^1\lambda_1   ) }du^3.
$$

$\mathbf{2})$

\begin{align}\label{53} ds^2=\frac{1}{ \Delta}[2du^0du^2+\frac{{du^1}^2}{\beta} - a_0 {du^2}^2 +{du^3}^2],\end{align}

  $$A_0=-\frac{1}{2\tilde{a}\beta} , \quad  A_1 =\frac{h}{\beta}, \quad A_2=h^2\tilde{a},\quad A_3=0. $$

  $$s_2 = \int(\frac{\lambda \psi + \lambda_2 - \beta{\lambda_1}^2 }{2(\lambda_0 -\frac{1}{2\tilde{a}\beta})})du^2,$$

$$
s^3 = {\varepsilon} \int\sqrt{(   \lambda f -\lambda_2  )}du^3.
$$

$\mathbf{3})$

\begin{align}\label{54} ds^2=\frac{1}{ \Delta}[2du^0du^2+\frac{(du^1-a_1 du^2)^2}{a+\beta} - a_0 {du^2}^2 +{du^3}^2],\end{align}

$$ A_0=0 \quad  A_1 =0, \quad A_2=h,\quad A_3=0. $$

 $$s_2 = \int(\frac{\lambda \psi + \lambda_2 - (\beta{\lambda_1}^2)}{2\lambda_0})du^2,$$

$$
s^3 = {\varepsilon} \int\sqrt{( \lambda f -(\lambda_2 + a^{pq}\lambda_p \lambda_q +2(h^0\lambda_0)   )}du^3.
$$

$\mathbf{4})$

\begin{align}\label{55} ds^2=\frac{1}{ \Delta}[2du^0du^2+\frac{(du^1-a_1 du^2)^2}{\beta} - a_0 {du^2}^2 +{du^3}^2],\end{align}

  $$A_0=0, \quad  A_1 =\frac{\gamma}{\beta}, \quad A_2=h + a_1 \frac{\gamma}{\beta} A_1,\quad A_3=0.$$

  $$s_2 = \int(\frac{\lambda \psi + \lambda_2 - (\beta{\lambda_1}^2 +2\gamma \lambda_1)}{2\lambda_0 })du^2,$$

$$
s^3 = {\varepsilon} \int\sqrt{(\lambda f -(\lambda_2 + a^{pq}\lambda_p \lambda_q +2h^0\lambda_0)} du^3.
$$

$\mathbf{5})$
\begin{align}\label{56} ds^2=\frac{1}{ \Delta}(\frac{2du^2(b_1du^0 - b_0 du^1)} {(b_1+\phi b^0)} +\frac{(\phi du^0+du^1)^2}{(b_1+\phi b^0)^2 +{du^3}^2 +{du^3}^2}+{du^3}^2) ,\end{align}

  $$A_0=\frac{\phi h +b_1 \sigma}{(b_1+\phi b^0)}, \quad  A_1 =\frac{h - a_0\sigma}{(b_1+\phi b^0)}, \quad A_2=0,\quad A_3=0. $$

 $$s_2 = \int\frac{\lambda \psi + \lambda_2 }{2(\lambda_0 - \phi \lambda_1 +\sigma)}du^2,$$

$$
s_3 = {\varepsilon} \int\sqrt{(   \lambda f -\lambda_2 -(\lambda_p b^p-h)^2 )}du^3.
$$

$\mathbf{6})$

\begin{align}\label{57} ds^2=\frac{1}{ \Delta}[2du^0du^2+\frac{(\phi du^0+du^1)^2}{\beta} +{du^3}^2],\end{align}

  $$A_0= \sigma, \quad  A_1 =\frac{\gamma }{\beta}, \quad A_2=0,\quad A_3=0. $$

 $$s_2 = \int\frac{\lambda \psi + \lambda_2 - (\beta{\lambda_1}^2 +2\gamma \lambda_1)}{2(\lambda_0 - \phi \lambda_1 +\sigma)}du^2,$$

$$
s_3 = {\varepsilon} \int\sqrt{(   \lambda f -\lambda_2  )}du^3.
$$

$\mathbf{7})$

\begin{align}\label{58} ds^2=\frac{1}{ \Delta}[2du^0du^2+\frac{{du^1}^2}{a} - \frac{\tilde{a}}{a} {du^2}^2 +{du^3}^2],\end{align}

 $$A_0=\sigma, \quad  A_1 =\frac{\sqrt{\tilde{a}\sigma^2+\tilde{b}}}{a}, \quad A_2=0,\quad A_3=0. $$

 $$s_2 = \int\frac{\lambda \psi + \lambda_2 - 2\sqrt{\tilde{a}\sigma^2+\tilde{b}}\lambda_1}{2(\lambda_0 +\sigma)}du^2,$$

 $$
s^3 = {\varepsilon_3} \int\sqrt{\varepsilon(   \lambda f -(\lambda_2 + \frac{\tilde{a}}{a}{\lambda_0}^2 +a{\lambda_1}^2)}du^3.
$$

$\mathbf{8})$

\begin{align}\label{59} ds^2=\frac{1}{ \Delta}[2du^0du^2+(du^1-adu^2)^2 - a^2 {du^2}^2 +{du^3}^2],\end{align}

 $$A_0=\sigma, \quad  A_1 =(\tilde{c}/(\sigma-\tilde{e})+(\tilde{e}-\sigma)a, \quad A_2= \frac{-a\tilde{e}}{\sigma-\tilde{e}}.$$

 $$s_2 = \int(\frac{\lambda \psi + \lambda_2 - (2(\tilde{c}/(\sigma-\tilde{e})+\tilde{e}a)\lambda_1)- (\tilde{c}/(\sigma+\tilde{e}))^2}{2(\lambda_0  +\sigma)})du^2,$$

$$
s^3 = {\varepsilon} \int\sqrt{(\lambda f -(\lambda_2 + (a\lambda_0 + \lambda_1)^2 +2(\tilde{e}a^2\lambda_0+ \tilde{e}a\lambda_1))-(\tilde{e}a)^2+2\tilde{c}a}du^3.
$$

$\mathbf{9})$

\begin{align}\label{60} ds^2=\frac{1}{ \Delta}[\frac{(du^0+du^1/{\phi)^2} +2du^2(\beta du^0 - a du^1{/\phi}) -a\beta{du^2}^2}{a+\beta} +{du^3}^2],\end{align}

  $$A_0=0, \quad  A_1 =\frac{-\sigma}{\phi}, \quad A_2=0,\quad A_3=0.$$

 $$s_2 = \int(\frac{\lambda \psi + \lambda_2 - (\beta{\phi}^2{\lambda_1}^2 -2\lambda_1 \beta \phi \sigma)+\beta{\sigma}^2}{2(\lambda_0 - \phi \lambda_1 +\sigma)})du^2,$$

$$
s^3 = {\varepsilon} \int\sqrt{(\lambda f -(\lambda_2 + a{\lambda_1}^2 )}du^3
$$

$\mathbf{10})$

\begin{align}\label{61} ds^2=\frac{1}{ \Delta}[2du^0du^2+\frac{(\phi du^0+du^1)^2}{a+\beta}+{du^3}^2],\end{align}

 $$A_0=\sigma, \quad  A_1 =0, \quad A_2=0,\quad A_3=0.$$

 $$s_2 = \int(\frac{\lambda \psi + \lambda_2 - \beta{\lambda_1}^2}{2(\lambda_0 - \phi \lambda_1 +\sigma)})du^2,$$

$$
s^3 = {\varepsilon} \int\sqrt{\lambda f -(\lambda_2 + a{\lambda_1}^2  )}du^3.
$$

Acknowledgments {This work was supported by the Ministry of Science and Higher Education of the Russian Federation, project FEWF-2020-0003.}.

\end{document}